\def\simlt{\hbox{ \rlap{\raise 0.425ex\hbox{$<$}}\lower 0.65ex\hbox{$\sim$} }}
\def\simgt{\hbox{ \rlap{\raise 0.425ex\hbox{$>$}}\lower 0.65ex\hbox{$\sim$} }}
\def\that{{\hat t}}
\def\msun{ {\rm M_\odot} }
\def\ee#1{\times 10^{#1}}

\documentstyle[aaspp4,epsf,astrobib]{article}
\begin{document}

\bibliographystyle{apj}

\title{EROS and MACHO Combined Limits on Planetary Mass Dark Matter
in the Galactic Halo}

\author{
          C.~Alcock\altaffilmark{1,2}, 
        R.A.~Allsman\altaffilmark{3},
          D.~Alves\altaffilmark{1,4},
          R.~Ansari\altaffilmark{5},
        \'E.~Aubourg\altaffilmark{6},
        T.S.~Axelrod\altaffilmark{7},
          P.~Bareyre\altaffilmark{6},
      J.-Ph.~Beaulieu\altaffilmark{8,9},
        A.C.~Becker\altaffilmark{2,10},
        D.P.~Bennett\altaffilmark{1,2,11},
          S.~Brehin\altaffilmark{6},
          F.~Cavalier\altaffilmark{5},
          S.~Char\altaffilmark{12},
        K.H.~Cook\altaffilmark{1,2},
          R.~Ferlet\altaffilmark{8},
          J.~Fernandez\altaffilmark{12},
        K.C.~Freeman\altaffilmark{7},
          K.~Griest\altaffilmark{2,13},
         Ph.~Grison\altaffilmark{8},
          M.~Gros\altaffilmark{6},
          C.~Gry\altaffilmark{14},
          J.~Guibert\altaffilmark{15},
          M.~Lachi\`eze-Rey\altaffilmark{6},
          B.~Laurent\altaffilmark{6},
        M.J.~Lehner\altaffilmark{16},
        \'E.~Lesquoy\altaffilmark{6},
          C.~Magneville\altaffilmark{6},
        S.L.~Marshall\altaffilmark{1,2},
        \'E.~Maurice\altaffilmark{17},
          A.~Milsztajn\altaffilmark{6},
          D.~Minniti\altaffilmark{1},
          M.~Moniez\altaffilmark{5},
          O.~Moreau\altaffilmark{15},
          L.~Moscoso\altaffilmark{6},
          N.~Palanque-Delabrouille\altaffilmark{6},
        B.A.~Peterson\altaffilmark{7},
        M.R.~Pratt\altaffilmark{18},
          L.~Pr\'ev\^ot\altaffilmark{17},
          F.~Queinnec\altaffilmark{6},
        P.J.~Quinn\altaffilmark{19},
          C.~Renault\altaffilmark{6,20},
          J.~Rich\altaffilmark{6},
          M.~Spiro\altaffilmark{6},
        C.W.~Stubbs\altaffilmark{2,7,10},
          W.~Sutherland\altaffilmark{21},
          A.~Tomaney\altaffilmark{2,10},
          T.~Vandehei\altaffilmark{2,13},
          A.~Vidal-Madjar\altaffilmark{8},
          L.~Vigroux\altaffilmark{6},
          S.~Zylberajch\altaffilmark{6}
        }


\altaffiltext{1}{Lawrence Livermore National Laboratory, Livermore, CA 94550}

\altaffiltext{2}{Center for Particle Astrophysics,
        University of California, Berkeley, CA 94720}

\altaffiltext{3}{Supercomputing Facility, Australian National University,
        Canberra, ACT 0200, Australia}

\altaffiltext{4}{Department of Physics, University of California,
        Davis, CA 95616 }

\altaffiltext{5}{Laboratoire de l'Acc\'el\'erateur Lin\'eaire,
        IN2P3 CNRS, Universit\'e Paris-Sud, F-91405 Orsay Cedex, France}

\altaffiltext{6}{CEA, DSM, DAPNIA, Centre d'\'Etudes de Saclay,
        F-91191 Gif-sur-Yvette Cedex, France}

\altaffiltext{7}{Mt.~Stromlo and Siding Spring Observatories,
        Australian National University, Weston, ACT 2611, Australia}

\altaffiltext{8}{Institut d'Astrophysique de Paris, INSU CNRS, 
        98 bis Boulevard Arago, F-75014 Paris, France}

\altaffiltext{9}{Kapleyn Astronomical Institute, 
        N-9700 AV Groningen, The Netherlands}

\altaffiltext{10}{Departments of Astronomy and Physics,
        University of Washington, Seattle, WA 98195}

\altaffiltext{11}{Physics Department, University of Notre Dame,
        Notre Dame, IN 46556}

\altaffiltext{12}{Universidad de la Serena, Faculdad de Ciencias,
        Departemento de Fisica, Casilla 554, La Serena, Chile}

\altaffiltext{13}{Department of Physics, University of California,
        San Diego, CA 92093}

\altaffiltext{14}{Laboratoire d'Astronomie Spatiale de Marseille,
        Traverse du Siphon, Les Trois Lucs, F-13120, Marseille, France}

\altaffiltext{15}{Centre d'Analyse des Images de l'INSU, Observatoire de Paris,
        61 avenue de l'Observatoire, F-75014, France}

\altaffiltext{16}{Department of Physics, University of Sheffield,
        Sheffield S3 7RH, UK}

\altaffiltext{17}{Observatoire de Marseille, 2 place Le Verrier,
        F-13248 Marseille Cedex 04, France}

\altaffiltext{18}{Center for Space Research, MIT, Cambridge, MA 02139}

\altaffiltext{19}{European Southern Observatory, Karl Schwarzschild Str. 2,
        D-85748 G\"arching bei M\"unchen, Germany}

\altaffiltext{20}{Max-Planck-Institut f\"ur Kernphysik, Postfach 10 39 80,
        D-69029 Heidelberg, Germany}

\altaffiltext{21}{Department of Physics, University of Oxford,
        Oxford OX1 3RH, UK}

\vspace{2in}

\begin{abstract} 
The EROS and MACHO collaborations have each published upper limits on the
amount of planetary mass dark matter in the Galactic Halo obtained from
gravitational microlensing searches. In this paper the two limits are
combined to give a much stronger constraint
on the abundance of low mass MACHOs.
Specifically, objects with masses
$10^{-7}~\msun \simlt m \simlt 10^{-3}~\msun$
make up less than $25$\% of the halo dark matter for most models considered,
and less than $10$\% of a standard spherical halo is made of MACHOs
in the $3.5\ee{-7}~\msun < m < 4.5\ee{-5}~\msun$
mass range.
\end{abstract}

\twocolumn

If a significant fraction of the dark matter in the Galactic Halo
is in the form of MACHOs (objects of masses $m \simgt 10^{-8}~\msun$),
then these objects can be detected via gravitational microlensing
\cite{pac86}, which is the temporary brightening of a background star
as the unseen object passes close to the line of sight.
The EROS and MACHO collaborations have been monitoring the brightnesses of
millions of stars in the Magellanic Clouds for several years
to search for these gravitational microlensing events,
and several candidate events have been detected \cite{lmc2,ansari},
with Einstein ring diameter crossing times
$33~{\rm days} < \that < 266~{\rm days}$.
For a MACHO of mass $m$, the average timescale of a microlensing event
(assuming a standard spherical halo) is given by \cite{griest91a}
\begin{equation}
\langle\that\rangle \sim 130 \sqrt{m/\msun}~{\rm days},
\end{equation}
so these events correspond to lens masses $m \simgt 0.1~\msun$.
For planetary mass objects
($10^{-8}~\msun \simlt m \simlt 10^{-3}~\msun$), the event durations
become quite short, from a fraction of an hour to a few days.
Both groups have reported upper limits on
the abundance of planetary mass dark matter
\cite{spikelmc2,eros3}, but because there is only a small overlap
in exposure between the projects,
it is possible to produce stronger limits
by combining the largely independent results of the two groups.

The EROS search for low mass MACHOs
(a part of the first phase of the EROS experiment)
is described in
detail in \citeN{eros3} and
\citeN{eros2}.
The EROS program (Exp\'{e}rience de Recherche d'Objets Sombres)
used a CCD camera  at the European Southern Observatory at La Silla, Chile,
devoted to the detection of events with small durations
occurring between 1991 and 1995. 
One field of about half a square degree was observed
about 20 times per night in two colors and
 contains about 150,000 stars.
The first three years were devoted to the observation of one field in the bar
of the LMC, the last year to one field in the center of
the SMC. Each year of data was analyzed seperately.
The search is sensitive to
microlensing durations ranging from 15 minutes to a few
days on stars brighter 
than about 19.5 magnitude in V band.
More than 19,000 images have been
processed using custom designed fast photometric
reconstruction software to produce light curves.
None of the 350,000 good light curves exhibits a form which is consistent with
a microlensing event.
Using the detection efficiency, largely affected by blending effects
and the finite size of the
source at the
lowest durations, objects in
the range $2\ee{-7}~\msun < m < 2\ee{-3}~\msun$
can be excluded 
as a major constituent of the dark halo for different models of the Galaxy.

The MACHO collaboration search for low mass MACHOs is described in detail
in \citeN{spikelmc2} and \citeN{spikelmc2proc}. In summary, because the 
initial observing strategy of the MACHO collaboration was designed to
maximise the detection rate for lenses in the brown-dwarf range
$10^{-3}$ to $0.1~\msun$
(corresponding to event durations
of a few days or longer), images of a given field were taken at most once
or twice per clear night. For planetary mass lenses whose event durations
are typically less than one day, there would be at most one or two (if any)
magnified points on the lightcurve. If such an event were found it could
not be classified as microlensing, but strong limits can be placed on the
amount of dark matter in the form of low mass MACHOs if few of these events
are found. Analysis of two years of data (from 20 July 1992 through
26 October 1994) on 8.6 million stars in 22 LMC fields found none of these
``spike'' events, and it was reported that MACHOs in the mass range
$2.5\ee{-7}~\msun < m < 5.2\ee{-4}~\msun$ can not make up the entire
mass of a standard spherical halo.

Even though the two experiments use very different analysis techniques,
they produce fairly similar results. This is because the MACHO analysis
has a very low efficiency ($\sim 1$\% for magnifications greater
than 1.042)
but a very large number of stars, and the EROS
analysis gives a fairly high efficiency 
(up to $40$\% for a magnification greater than 1.08)
on a small number of stars.
Therefore there is little overlap in exposure
for the two projects, and the two limits can be combined after
taking this into account.

\begin{figure}
\plotone{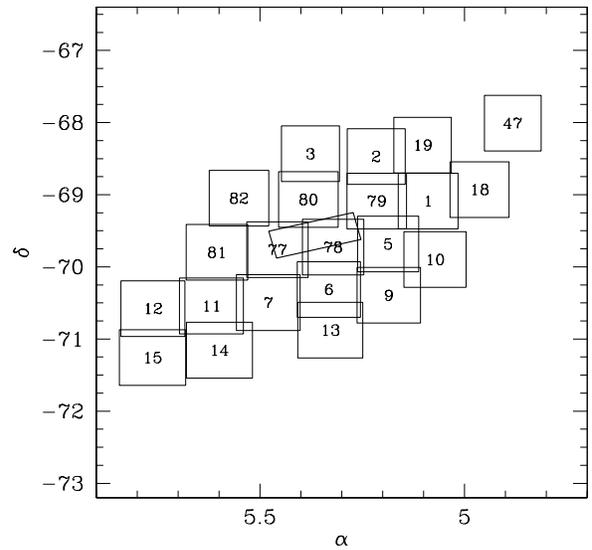}
\caption{The locations (J2000)
of the MACHO fields used in this analysis and the
location of the EROS field (thick line).}
\label{fig:fields}
\end{figure}

The 22 MACHO LMC fields used in this analysis are shown in
Figure~\ref{fig:fields}, along with the field used by the
EROS experiment. The redundant measurements
were eliminated by removing those stars in the MACHO database
which lie in the EROS field on nights when those stars were imaged
by both collaborations.
(The MACHO data were removed because the efficiency is much lower so
less sensitivity was lost.) This reduced the
MACHO effective exposure by about 10\%.
The MACHO limits were then recalculated,
and the number of expected events were simply added to the number of
expected events from the EROS analysis.

Each collaboration has used different halo models when reporting their
results, but once the detection efficiency is known
it is fairly simple to calculate the combined limits for both
sets of models, which are summarized in Table~1. Models 1 - 5 are used by
the EROS collaboration, and models S - G are used by the MACHO collaboration.
Models 1 - 4 and A - G are the power-law models of \citeN{evans94b},
and models 5 and S are simple spherical models as described in
\citeN{griest91a} and \citeN{ansari}.
With these 13~models, we cover a fairly large range of
reasonable Galactic Halo
mass and velocity distributions.
(Model E has the bulk of the Galactic mass in the disk. This
is very likely an unreasonable assumption, but we include this model
anyway for comparison to previous publications.)

\begin{figure}
\plotone{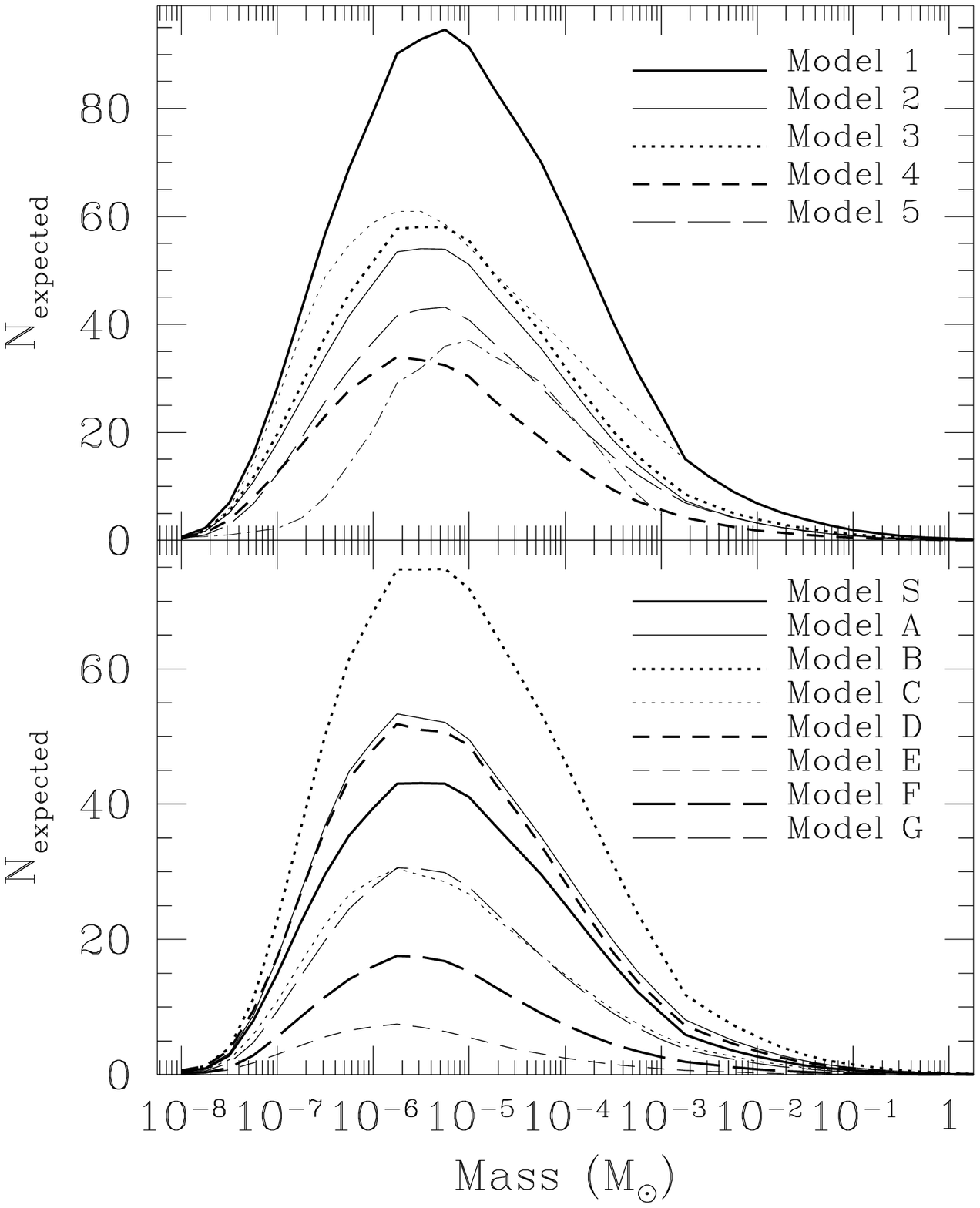}
\caption{Total number of expected events versus lens mass for the
combined MACHO and EROS results.
The five EROS models are shown on
the top plot and the eight MACHO models are shown on the bottom.
Also shown in the top plot is the contribution to the results for
model 1 from the EROS results (thin dotted line) and the MACHO results
(dot-dash line). The relative contributions are roughly the same for
all models.}
\label{fig:nexp2}
\end{figure}

\begin{figure}
\plotone{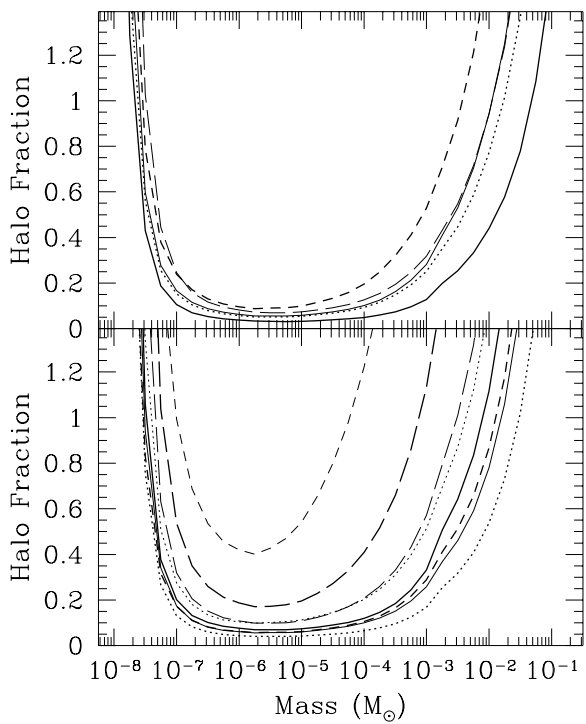}
\caption{Halo fraction upper limit (95\% c.l.)
versus lens mass for the five EROS models
(top) and the eight MACHO models (bottom). The line coding is the same
as in Figure~2.}
\label{fig:frac2}
\end{figure}

\begin{figure}
\plotone{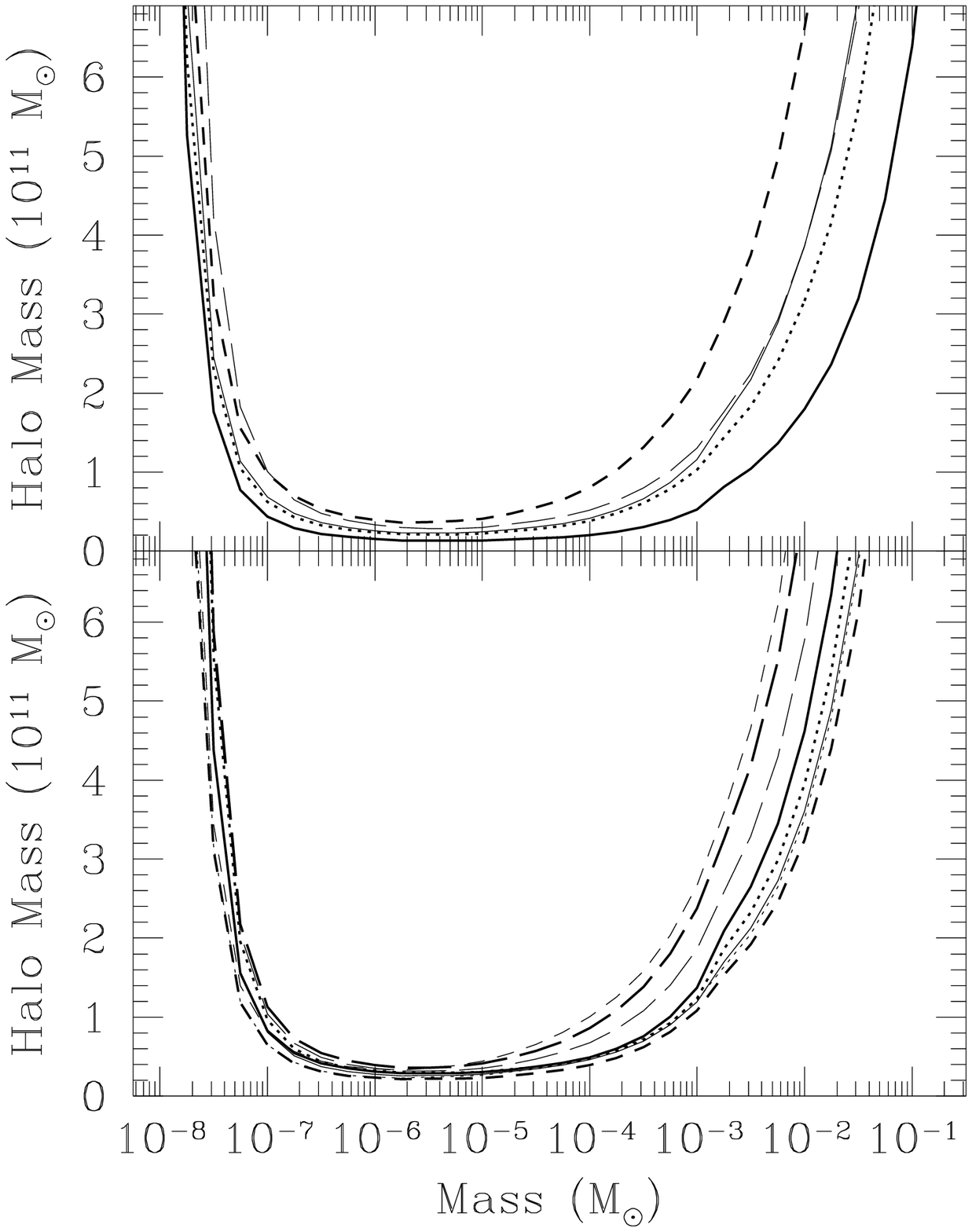}
\caption{Upper limit (95\% c.l.)
on total halo mass in MACHOs
versus lens mass for the five EROS models
(top) and the eight MACHO models (bottom). The line coding is the same
as in Figure~2.}
\label{fig:mass2}
\end{figure}

The number of expected events as a function of lens mass
(assuming a $\delta$-function mass distribution) can be found in
Figure~\ref{fig:nexp2} for the five EROS models and
the eight MACHO models.
Also, Figure~\ref{fig:frac2}
shows the $95$\% confidence level
upper limit on the fraction of the halo dark matter which can
consist of MACHOs of mass $m$. Here it can be seen that for most
models, objects with masses $10^{-7}~\msun \simlt m \simlt 10^{-3}~\msun$
comprise less than $25$\% of the halo dark matter,
and less than $10$\% of a standard spherical halo
(model 5) is made of MACHOs
in the $3.5\ee{-7}~\msun < m < 4.5\ee{-5}~\msun$
mass range.
Because we are using $\delta$-function mass distributions, 
any mass function consisting entirely of masses
in the excluded intervals is also eliminated
\cite{griest91a}.
Figure~\ref{fig:mass2} shows the amount of halo mass that can be
comprised of MACHOs of mass $m$, which is a more model-independent
limit \cite{spikelmc2}. Here it is shown that for all models considered,
a canonical halo mass of $4.1\ee{11}~\msun$ can not be
comprised entirely of MACHOs in the mass range
$10^{-7}~\msun \simlt m \simlt 10^{-3}~\msun$, and MACHOs with masses
$10^{-7}~\msun \simlt m \simlt 10^{-4}~\msun$ account for less
than $1\ee{11}~\msun$ of the total halo mass.  
These are the strongest limits published on
dark matter in this mass range, and the limits will get stronger in time
as more data are collected and
analyzed. 
The MACHO collaboration is currently analyzing
two more years of LMC data,
and is also continuing to collect data. However,
the EROS short duration microlensing search
was discontinued in April 1995,
and because of the temporal sampling of the new data now
being collected \cite{eros4}, a spike analysis will
not be performed.

\acknowledgements

The EROS collaboration is grateful for the support given to the
project by the technical staff at
ESO La Silla.
The MACHO collaboration is
grateful for the support given the project by the technical
staff at the Mt. Stromlo Observatory.  Work performed at LLNL is
supported by the DOE under contract W-7405-ENG-48.  Work performed by the
Center for Particle Astrophysics personnel is supported by the NSF
through AST 9120005.  The work at MSSSO is supported by the Australian
Department of Industry, Science, and Technology. K.G. acknowledges
support from DOE, Alfred P. Sloan, and Cotrell Scholar awards.  
C.S. acknowledges the generous support of the Packard and Sloan Foundations.
W.S. is supported by a PPARC Advanced fellowship.
M.J.L. acknowledges support from an IGPP mini-grant.

\bibliography{comb_lim}

\begin{thebibliography}{}

\bibitem[\protect\citeauthoryear{Alcock et~al.}{Alcock
  et~al.}{1996}]{spikelmc2}
Alcock, C.,  et~al. 1996, \apj, 471, 774

\bibitem[\protect\citeauthoryear{Alcock et~al.}{Alcock et~al.}{1997}]{lmc2}
Alcock, C.,  et~al. 1997, \apj, 486, 697

\bibitem[\protect\citeauthoryear{Ansari et~al.}{Ansari et~al.}{1996}]{ansari}
Ansari, R.,  et~al. 1996, \aap, 314, 94

\bibitem[\protect\citeauthoryear{Evans}{Evans}{1994}]{evans94b}
Evans, N.~W. 1994, \mnras, 267, 333

\bibitem[\protect\citeauthoryear{Griest}{Griest}{1991}]{griest91a}
Griest, K. 1991, \apj, 366, 412

\bibitem[\protect\citeauthoryear{Lehner et~al.}{Lehner
  et~al.}{1996}]{spikelmc2proc}
Lehner, M.~J.,  et~al. 1996, in {Dark Matter in the Universe}, ed. D.~B. Cline

\bibitem[\protect\citeauthoryear{Paczy{\'n}ski}{Paczy{\'n}ski}{1986}]{pac86}
Paczy{\'n}ski, B. 1986, \apj, 304, 1

\bibitem[\protect\citeauthoryear{Palanque-Delabrouille
  et~al.}{Palanque-Delabrouille et~al.}{1998}]{eros4}
Palanque-Delabrouille, N.,  et~al. 1998, \aap, in press

\bibitem[\protect\citeauthoryear{Renault et~al.}{Renault et~al.}{1997}]{eros2}
Renault, C.,  et~al. 1997, \aap, 324, 69L

\bibitem[\protect\citeauthoryear{Renault et~al.}{Renault et~al.}{1998}]{eros3}
Renault, C.,  et~al. 1998, \aap, 329, 522

\end{thebibliography}

\onecolumn

\begin{deluxetable}{crrccccc}
\tablecaption{Halo Model Parameters.
\label{tablepl}}
\tablehead{
\colhead{Model} & 
\colhead{$\beta$\tablenotemark{a}} & 
\colhead{$q$\tablenotemark{b}} & 
\colhead{$v_0\,\rm (km/sec)$\tablenotemark{c}} & 
\colhead{$R_{\rm c}\,\rm(kpc)$\tablenotemark{d}} & 
\colhead{$R_0\,\rm(kpc)$\tablenotemark{e}} &
\colhead{$M_{50}\,\rm(10^{11}\msun)$\tablenotemark{f}}}
\startdata
1 & 0.0 & 1.00 & 269 & 5.6 & 7.9 & 4.10 \nl
2 & 0.0 & 1.00 & 203 & 5.6 & 7.9 & 4.10 \nl
3 & 0.0 & 0.75 & 204 & 5.6 & 7.9 & 4.10 \nl
4 & 0.0 & 0.75 & 154 & 5.6 & 7.9 & 4.10 \nl
5 & -- & -- & 185 & 7.8 & 7.9 & 4.10 \nl
S & -- & -- & 220 & 5.0 & 8.5 & 4.13 \nl
A & 0.0 & 1.00 & 200 & 5.0 & 8.5 & 4.62 \nl
B & -0.2 & 1.00 & 200 & 5.0 & 8.5 & 7.34 \nl
C & 0.2 & 1.00 & 180 & 5.0 & 8.5 & 2.36 \nl
D & 0.0 & 0.71 & 200 & 5.0 & 8.5 & 3.74 \nl
E & 0.0 & 1.00 & 90 & 20.0 & 7.0 & 0.82 \nl
F & 0.0 & 1.00 & 150 & 25.0 & 7.9 & 2.10 \nl
G & 0.0 & 1.00 & 180 & 20.0 & 7.9 & 3.26 \nl
\enddata
\tablenotetext{a}{$\beta$ gives the shape of the rotational velocity curve
($v_{\rm circ} \propto R^{-\beta})$.}
\tablenotetext{b}{$q$ is the halo flattening parameter.
($q = 1$ gives spherical halo, $q = 0.7$ represents ellipticity of E6).}
\tablenotetext{c}{$v_0$ is a normalization velocity
(which corresponds to $v_{\rm circ}$ if $\beta = 0$).}
\tablenotetext{d}{$R_{\rm c}$ is the Galactic core radius.}
\tablenotetext{e}{$R_0$ is the radius of the solar orbit.}
\tablenotetext{f}{$M_{50}$ is the total mass
of halo dark matter interior to 50~kpc from the Galactic center.}
\end{deluxetable}

\end{document}